\documentclass[12pt]{article}
\usepackage{amsthm,amssymb,amsmath}
\usepackage{graphicx}
\usepackage{epsfig}
\begin{document}
\begin{titlepage}
\begin{center}
\textbf{\large Gamma rays from the annihilation of singlet scalar dark matter}\\[1cm]
Carlos E. Yaguna\\[3mm]
{\it Departamento de F\'{i}sica Te\'orica C-XI and Instituto de F\'{i}sica Te\'orica UAM-CSIC,\\[3mm]
Universidad Aut\'onoma de Madrid, Cantoblanco, E-28049 Madrid, Spain}
\end{center}
\begin{abstract}
We consider an extension of the Standard Model by a singlet scalar that accounts for the dark matter of the Universe. Within this model we  compute the expected gamma ray flux from the annihilation of dark matter particles in a consistent way. To do so, an updated analysis of the parameter space of the model is first presented. By enforcing the relic density constraint from the very beginning,  the viable parameter space gets  reduced to just two variables: the singlet mass and the higgs mass. Current direct detection constraints are then found to require a singlet mass larger than $50$ GeV. Finally,  we compute the gamma ray flux and annihilation cross section  and show that a large fraction of the viable parameter space lies within the sensitivity of Fermi-GLAST.
\end{abstract}
\end{titlepage}
\section{Motivation}
A simple extension of the Standard Model that can explain the dark matter is the addition of a real scalar singlet and an unbroken $Z_2$ symmetry under which the singlet is odd while all other fields are even. Such a singlet, which couples directly only to the higgs boson and to itself, may indeed have the right relic density --in the standard cosmological model-- to explain the observed dark matter abundance. Even though this singlet extension has been studied several times \cite{McDonald:1993ex, Burgess:2000yq, Davoudiasl:2004be, Barger:2007im, Dick:2008ah}, a complete and consistent computation of the expected gamma ray flux from the annihilation of singlet dark matter has yet to be published. In  this paper we will fill that void. First, an updated analysis of the parameter space of the model will be presented. Then, we compute the direct detection cross section and show that present constraints require a singlet mass larger than $50$ GeV. Finally, after  obtaining the gamma ray flux and total annihilation cross section, we show that over most of the viable parameter space the singlet scalar model of dark matter is \emph{detectable} by Fermi-GLAST \cite{Gehrels:1999ri}.

The observation of gamma rays originating in the annihilation of dark matter particles is one of the most promising avenues to  determine the nature of dark matter. The recently launched Fermi Gamma-ray Space Telescope (FGST), formerly GLAST \cite{Gehrels:1999ri}, will improve, with respect to its predecessor EGRET \cite{egret}, the sensitivity to gamma rays in the energy range between $20$ MeV and $10$ GeV by more than one order of magnitude. More importantly, FGST will extend the high energy range to about $300$ GeV, making it an ideal experiment to search for  gamma rays from WIMP dark matter annihilation. It is therefore critical to determine the expected gamma ray flux  within diverse scenarios accounting for the dark matter.

The lightest neutralino in supersymmetric models is by far the most common dark matter candidate examined in the literature.   Even if less compelling from a theoretical point of view, the singlet scalar considered in this paper offers an interesting alternative to neutralino dark matter. Depending only on two new parameters, the singlet scalar model is very predictive and could be easily falsified. Moreover, the dark matter candidate is a scalar instead of a Majorana fermion and its main annihilation channels do not coincide with those of the neutralino. Direct and indirect detection signals  are thus expected to be different. Finally, the disparity between the implications for colliders searches of these  two dark matter models could not be more marked. The singlet scalar model predicts the existence of \emph{one} additional degree of freedom rather than a full spectrum of superpartners. The LHC, therefore, may soon shed some light on the identity of the dark matter particle and, in particular, on its supersymmetric or non-supersymmetric nature. In the meantime, it is important not to restrict ourselves to supersymmetric candidates.

Over the years, several authors have studied the phenomenology of the singlet extension of the Standard Model. The singlet scalar as a dark matter candidate  was initially proposed by  McDonald \cite{McDonald:1993ex} and was subsequently analyzed by Burgess \emph{et. al} \cite{Burgess:2000yq}.  Recently, Barger \emph{et al.} \cite{Barger:2007im} considered models with and without the $Z_2$ symmetry and investigated their expected phenomenology at the LHC. Regarding dark matter, they computed the singlet relic density as well as its direct detection cross section. Later on, in \cite{Dick:2008ah}, the gamma ray flux was computed but only for two specific values of the singlet-higgs coupling.  In this paper we update and expand these previous results in several respects. First, we consider the full mass range, including $m_S<M_W$ and $m_S>m_h$, for the singlet scalar. Second, we use the precise determination of the dark matter density obtained by the WMAP experiment \cite{Dunkley:2008ie} as well as the accurate computation of the relic density by micrOMEGAs \cite{Belanger:2006is} --which includes \emph{all} tree-level annihilation processes-- to obtain the viable parameter space of the model.  Third, we compute the direct detection cross section and take  into account the constraints from current experiments. Finally, we use state of the art techniques, as implemented in micrOMEGAs \cite{Belanger:2006is}, to compute the expected gamma ray flux along the viable regions of the singlet model.

The paper is organized as follows. In the next section we will introduce the model Lagrangian and will identify the new parameters that it contains. In section \ref{sec:rd} the relic density is computed and used to obtain the viable parameter space. Direct detection rates are then calculated and compared with the sensitivity of present and planned experiments in section \ref{sec:dd}. Finally, in section \ref{sec:gf}, we  compute the expected gamma ray flux and annihilation cross section and show that Fermi-GLAST will probe a large fraction of the viable parameter space of the singlet scalar model.    

\section{The model}
The Lagrangian that describes the model with an additional  scalar singlet, S, is
\begin{equation}
\mathcal{L}= \mathcal{L}_{SM}+\frac 12 \partial_\mu S\partial^\mu S-\frac{m_0^2}{2}S^2-\frac{\lambda_S}{4}S^4-\lambda S^2 H^\dagger H\,,
\label{eq:la}
\end{equation}
where $\mathcal{L}_{SM}$ denotes the Standard Model Lagrangian and $H$ is the higgs doublet. This Lagrangian is the most general renormalizable one that is compatible with  the $SU(3)\times SU(2)\times U(1)$ gauge invariance and with the symmetry $S\to -S$. The scalar singlet extension of the standard  model, therefore, contains only $3$ new parameters: $m_0$, $\lambda$, and $\lambda_S$. Because it only determines the strength of the singlet self-interactions, $\lambda_S$ is  unconstrained and largely irrelevant to the phenomenology of the model. In the following we will simply require $\lambda_S\lesssim 1$ so as to guarantee a perturbative treatment. Notice, from \eqref{eq:la}, that the singlet couples to Standard Model fields only through the higgs boson and that such interaction is determined by the parameter $\lambda$.

A detailed analysis of the scalar potential was already presented in \cite{Burgess:2000yq}. Here, we briefly review, for completeness, the constraints that the potential must satisfy. In the unitary gauge, the scalar potential takes the form 
\begin{equation}
V=\frac{m_0^2}{2} S^2+\frac\lambda 2S^2h^2+\frac{\lambda_S}{4}S^4+\frac{\lambda_h}{4}(h^2-v_{EW}^2)^2
\end{equation}
where $v_{EW}=246$ GeV and $\lambda_h$ is the higgs quartic coupling. The configuration $S=0$ and $h\neq 0$ is a local minimum of $V$ provided that $v_{EW}^2>0$ and $m_0^2+\lambda v_{EW}^2 >0$. Another local minimum, with $h=0$ and $S^2=-m_0^2/\lambda$, will exist whenever $m_0^2<0$ and $-\lambda m_0^2>\lambda_S\lambda_h v_{EW}^2$. In such case, to ensure that the former is the potential's global minimum we must require that $0<-m_0^2<v_{EW}^2\sqrt{\lambda_h\lambda_S}$. Once these conditions are satisfied, the S-dependent part of the potential can be rewritten as
\begin{equation}
V=\frac 12 m_S^2 S^2+\frac{\lambda_S}{4} S^4+\lambda v_{EW}S^2 h+\frac{\lambda}{2}S^2 h^2
\end{equation}
where $m_S^2=m_0^2+\lambda v_{EW}^2$, and $h$ represents the physical higgs boson with mass $m_h^2=\lambda_h v_{EW}^2$. In the following we take $m_S$ and $\lambda$ as the free parameters of the singlet scalar model.

\section{The viable parameter space}
\label{sec:rd}

\begin{figure}[t]
\begin{center}
 \includegraphics[scale=0.4]{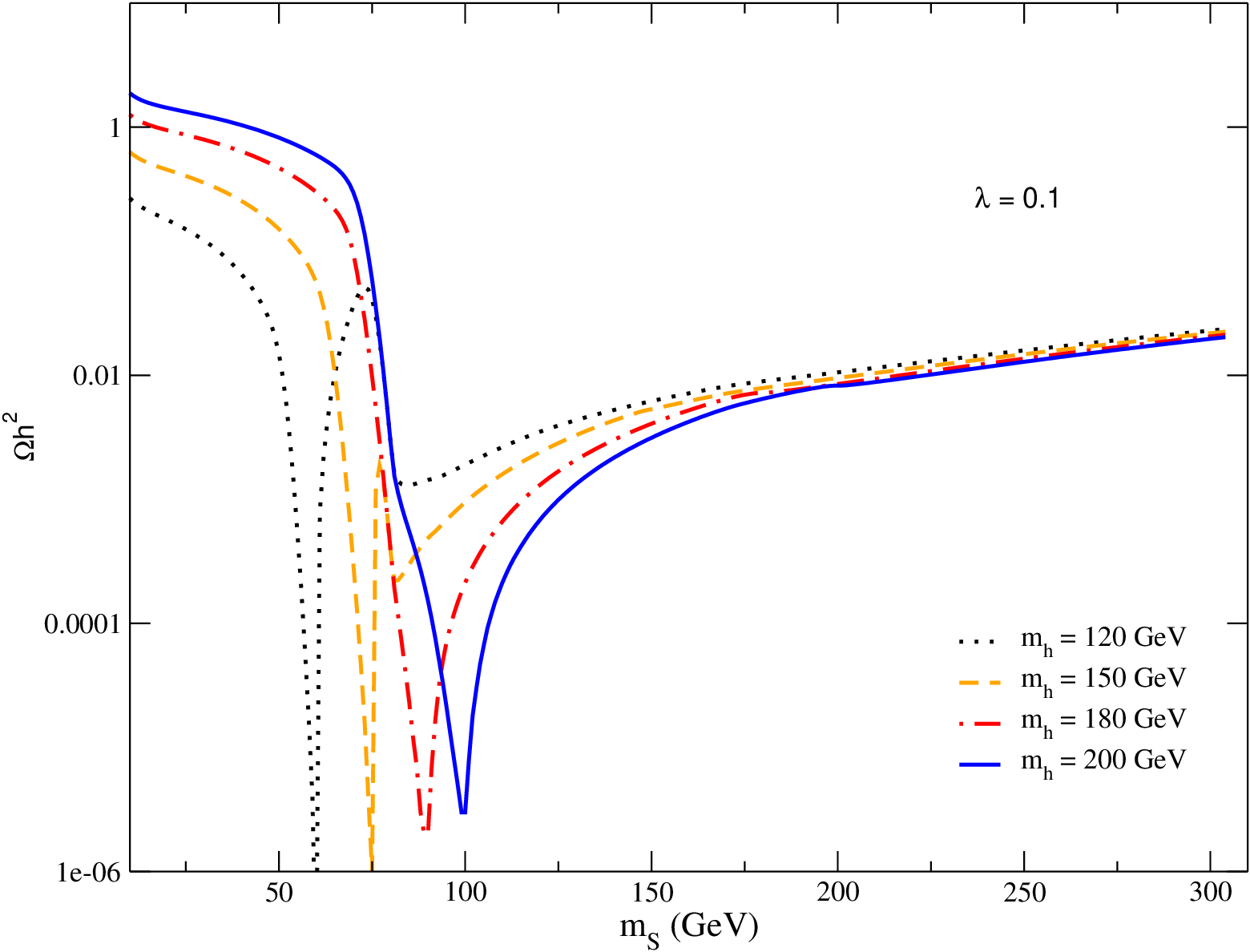}
\caption{The dark matter density as a function of $m_S$ for $\lambda=0.1$ and different values of the higgs mass.\label{rd}}
\end{center}

\end{figure}

In this section we compute the relic density of the singlet scalar and use it to impose the dark matter constraint, $\Omega_Sh^2=0.11$ \cite{Dunkley:2008ie}. From it, we obtain the viable parameter space of the model with a singlet scalar.

Singlets can annihilate through s-channel higgs boson exchange into a variety of final states: $f\bar f$, $W^+W^-$, $Z^0Z^0$, and $hh$. Additionally, they can also annihilate into $hh$ either directly or through singlet exchange. As a general rule, the final state $W^+W^-$ tends to dominate the total annihilation cross section whenever such channel is open. A light singlet,  $m_S<M_W$, will annihilate mainly into the $b\bar b$ final state. An intermediate mass singlet, $m_W<m_S<m_t$, annihilates mostly into $W^+W^-$, with additional contributions from  $Z^0Z^0$ and, if allowed, $hh$. For a heavier singlet, $m_S> m_t$, the pattern is similar, as the $t\bar t$ channel gives a non-negligible but subdominant contribution. To accurately compute $\Omega_Sh^2$ we use the micrOMEGAs package \cite{Belanger:2006is}, which can calculate the relic density in a generic dark matter model.

Figure \ref{rd} shows the relic density as a function of $m_S$ for $\lambda=0.1$ and different values of the higgs mass. Notice that the scalar singlet model can explain the dark matter naturally --that is, without any fine-tuning in the parameters. Indeed, for $\lambda=0.1$ and  $m_S$ around the electroweak scale, the predicted relic density lies in the correct range to be compatible with the observations. The most noticeable feature from this figure is the drastic suppression of the relic density that takes place at the higgs resonance. In fact, around $2m_S\sim m_h$ the relic density is orders of magnitude smaller than anywhere else. The effect of the $W^\pm$ threshold is also seen to be important. Above it, $m_S>m_W$, the $W^+W^-$ annihilation channel is open and consequently the relic density tends to be smaller than below it. 

\begin{figure}[t]
\begin{center}
 \includegraphics[scale=0.4]{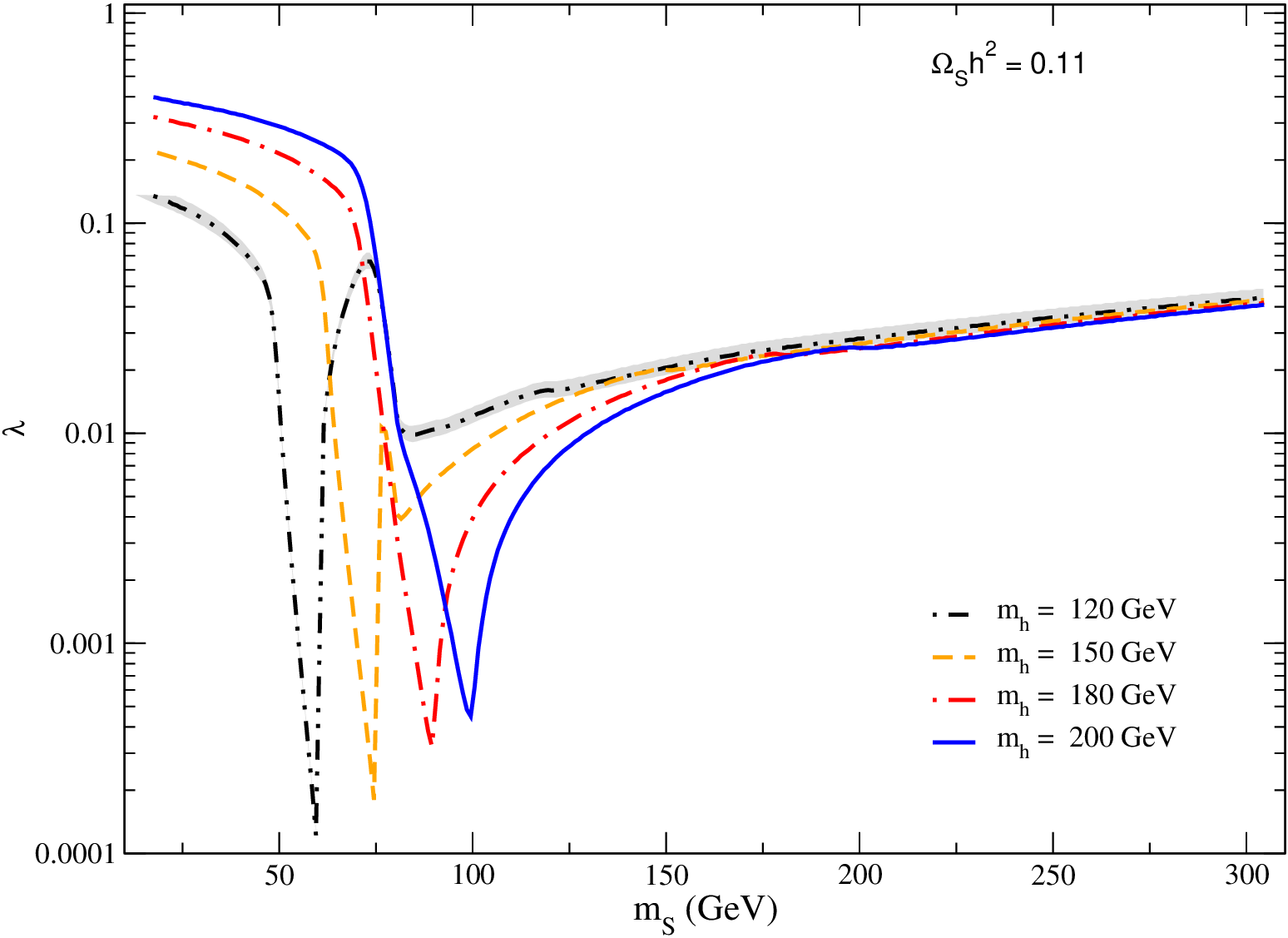}
\caption{The viable parameter space of the scalar singlet model. Along the lines the dark matter constrained is satisfied. We use $m_h=120,150,180, 200$ GeV as reference values for the higgs mass. The grey area surrounding the line corresponding to $m_h=120$ GeV shows the region compatible with the observed dark matter density at $2\sigma$.\label{pspace}}
\end{center}
\end{figure}

It is clearly seen from figure \ref{rd} that the value of the higgs mass, even if a standard model parameter,  is critical for the computation of the singlet relic density. From direct searches at LEP,  a lower limit on the higgs mass can be obtained, $m_h>114.4$ GeV \cite{Barate:2003sz}. When this bound is combined with electroweak precision measurements, an  upper limit --at the $95\%$ C. L.-- of $182$ GeV \cite{Alcaraz:2007ri} is derived. Throughout this paper we will use $m_h=120,150,180,200$ GeV as reference values for the higgs mass.

For any given pair ($m_h$,$m_S$) there exists a unique value of $\lambda$ such that  the dark matter constraint, $\Omega_S h^2 = 0.11$, is fulfilled.
By imposing the dark matter constraint, therefore, the variable $\lambda$ can be effectively eliminated for given values of  $m_h$ and $m_S$, reducing the viable parameter space to a two dimensional volume. In figure \ref{pspace} we show, in the plane ($\lambda$,$m_S$), lines that are compatible with the observed dark matter density for different values of $m_h$. Away from the higgs resonance, the typical value of $\lambda$ is $\mathcal{O}(10^{-1}-10^{-2})$. Light singlets, which annihilate mostly into $b\bar b$, require larger values of $\lambda$ to obtain the observed relic density. In contrast, at the higgs resonance, the annihilation tends to be  more efficient and much smaller values of $\lambda$ are needed to satisfy the dark matter constraint. To be concrete, we will not consider singlet masses above $300$ GeV, though such high values are also allowed. Figure \ref{pspace} defines what we call the \emph{viable} parameter space of the singlet scalar model. In the following we will compute the direct detection rates and the gamma ray flux only along these viable lines.

\section{Direct detection}
\label{sec:dd}

\begin{figure}[t]
\begin{center}
 \includegraphics[scale=0.4]{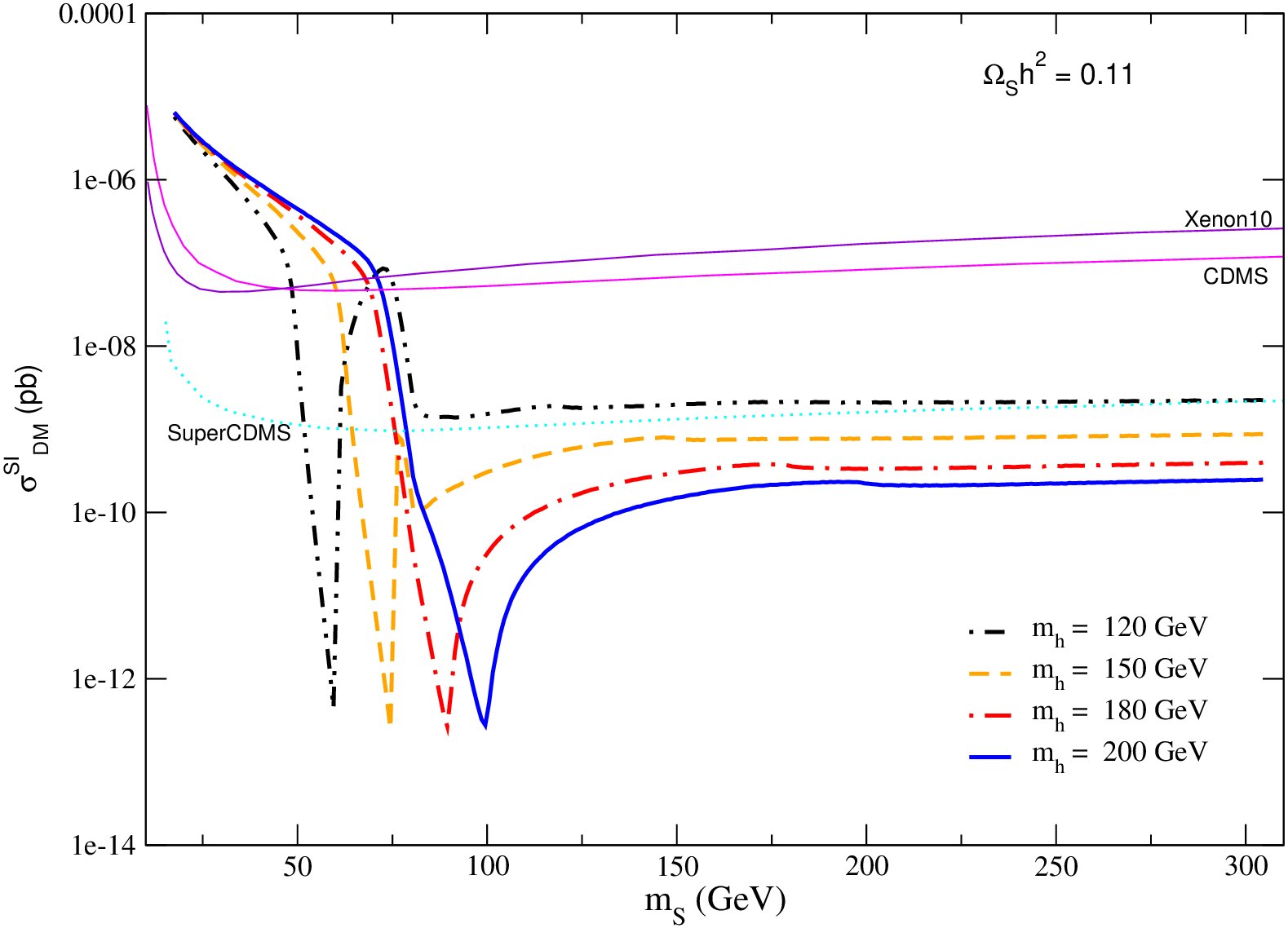}
\caption{The spin-independent proton-singlet cross section as a function of $m_S$ for different values of the higgs mass. The thin lines show the present constraint from XENON10 and CDMS. The dotted line corresponds to the expected sensitivity of SuperCDMS. Along the lines $\Omega_S h^2=0.11$.\label{sip}}
\end{center}
\end{figure}

Dark matter particles can scatter elastically on nuclei and be detected through nuclear recoil in direct detection experiments. In the non-relativistic limit, the dark matter-nucleon amplitude receives two contributions: the scalar or spin-independent interaction and the axial-vector or spin-dependent part. For the singlet scalar, the spin-dependent interaction vanishes so only the spin-independent part can give a signal. 
 
Figure \ref{sip} shows the spin-independent proton-singlet cross section as a function of the singlet scalar mass (see \cite{Belanger:2006is} to find out how this cross section is computed). Since the higgs-singlets coupling is small close to the higgs resonance, see figure \ref{pspace}, the cross section is highly suppressed in that region. A heavy singlet, $m_S\gtrsim 150$ GeV, has an interaction cross section around $10^{-9}\,\,\mathrm{pb}$.  Because the singlet interacts with nucleons via $t$-channel higgs exchange, the cross section typically decreases with the higgs mass, as observed in the figure. For reference,  current constraints from Xenon10 \cite{Angle:2007uj} and CDMS \cite{Ahmed:2008eu} are also displayed. They rule out singlet masses below $50$ GeV independently of the higgs mass. Nevertheless,  the possibility of explaining the DAMA signal \cite{Bernabei:2008yi} with a light singlet, $m_S\sim 5$ GeV, was recently explored in \cite{Andreas:2008xy}. Finally, notice  from the figure that future experiments, such as SuperCDMS \cite{Schnee:2005pj}, will probe a significant region of the viable parameter space.

\section{The gamma ray flux}
\label{sec:gf}

The role of the indirect detection of dark matter --that is, the detection of dark matter annihilation products-- is complementary to that of direct detection searches and will be crucial  in future dark matter studies. In principle, dark matter annihilations could be observed through gamma rays, neutrinos, or antimatter. Among them, the simplest and more robust is the gamma ray signal.

The gamma ray flux above some energy threshold $E_{thr}$ from a direction forming an angle $\psi$ with respect to the galactic center can be expressed as 

\begin{align}
\Phi(E_{thr})= &0.94\times 10^{-13}\mathrm{cm^{-2}s^{-1}}\nonumber\\
&\times \sum_i\int_{E_{thr}}^{m_S} dE_\gamma\frac{dN_\gamma^i}{dE_\gamma} \left(\frac{\langle\sigma_i v\rangle}{10^{-29} cm^3 s^{-1}}\right)\left(\frac{100 \mathrm{GeV}}{m_S}\right)^2 \bar J(\Delta\Omega) \Delta\Omega
\label{eqflux}
\end{align}
where $J(\psi)$ is the dimensionless line of sight integral around the direction $\psi$,
\begin{equation}
J(\psi)=\frac{1}{8.5\mathrm{kpc}}\left(\frac{1}{0.3 \mathrm{GeV cm^{-3}}}\right)^2\int\rho^2(l)dl(\psi)
\end{equation}
and  $\bar J(\Delta\Omega)$ is the average of $J(\psi)$ over the spherical region of  solid angle $\Delta\Omega$,
\begin{equation}
\bar J(\Delta\Omega)=\frac{1}{\Delta\Omega}\int_{\Delta\Omega}J(\psi)d\Omega
\end{equation}
From this expression we see that the gamma ray flux depends not only on particle physics parameters, such as $\langle\sigma v\rangle$ and $m_S$, but also on the unknown distribution of the dark matter, parameterized by $J(\Delta\Omega)$. 

\begin{figure}[t]
\begin{center}
 \includegraphics[scale=0.4]{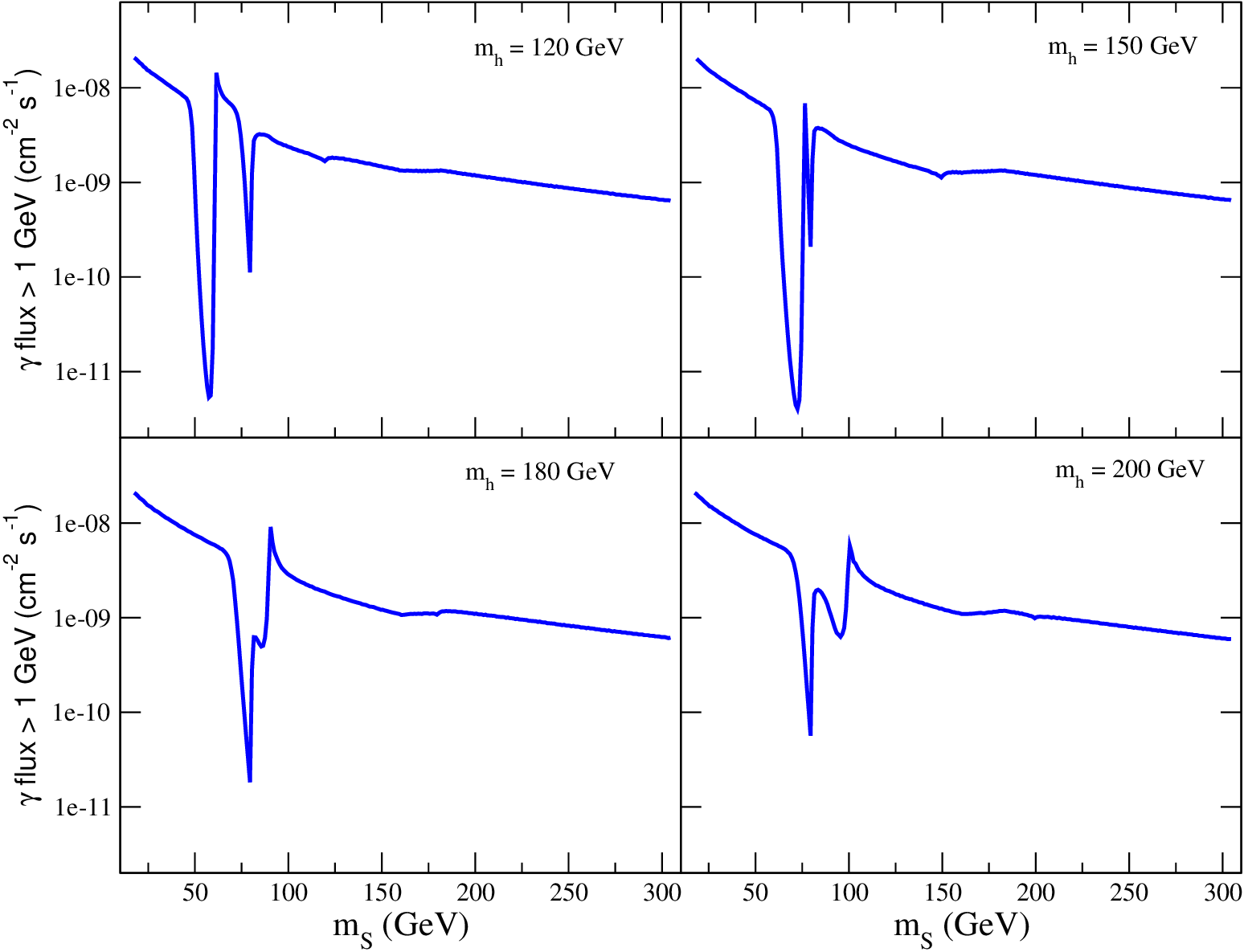}
\caption{The integrated photon flux above $1$ GeV as a function of the singlet scalar mass. The predicted gamma flux is from a $\Delta\Omega=10^{-3}$ sr region around the galactic center for a NFW profile.\label{flux}}
\end{center}
\end{figure}

In micrOMEGAs the photon spectrum from dark matter annihilations is obtained with a procedure similar to that implemented in DarkSUSY \cite{Gondolo:2004sc}. Such procedure relies on the use of tables of $\gamma$ production --obtained from PYTHIA \cite{Sjostrand:2006za}-- for each of the different \emph{basic} channels: $q\bar q$, $\mu^+\mu^-$, $\tau^+\tau^-$, $W^+W^-$ and $ZZ$. Concerning the distribution of dark matter,  in our analysis we take a NFW profile \cite{Navarro:1995iw}, which has a $\rho(r)\propto r^{-1}$ behaviour at small $r$, as the canonical example. For comparison we show, in table \ref{js}, the value of $\bar J(\Delta\Omega)$ for three different halo profiles --including NFW-- and two typical values of $\Delta\Omega$. 

\begin{table}[b]
\begin{center}
\begin{tabular}{lcc}
Profile & $\bar J(\Delta\Omega=10^{-3}$~sr) & $\bar J(\Delta\Omega=10^{-5}$~sr)\\
\hline 
NFW & $1.21\cdot 10^3$ & $1.26\cdot 10^4$ \\ 
Moore & $1.05\cdot 10^5$ & $9.46\cdot 10^6$ \\
Modified isothermal & $3.03\cdot 10^1$ & $3.03\cdot 10^1$ \\        
\hline
\end{tabular}
\caption{Values of $\left< J(0)\right>_{\Delta\Omega}$ for two different 
$\Delta\Omega$'s and for three different density profiles. See \cite{Cesarini:2003nr} for details.\label{js}}
\end{center}
\end{table}

Figure \ref{flux} shows the integrated gamma ray flux from the Galactic Center as a function of $m_S$ for different values of $m_h$. All these models are compatible with the EGRET constraint \cite{Cesarini:2003nr,Hunger:1997we}.  Notice that the flux has a dip not only at the higgs resonance but also at the $W$ threshold. This latter feature is due to the effect known as annihilation into forbidden channels \cite{Griest:1990kh}. The $1/m_S^2$ dependence, from (\ref{eqflux}),  is clearly visible in the figure. Lighter singlets generically yield a larger $\gamma$ ray flux.

\begin{figure}[t]
\begin{center}
 \includegraphics[scale=0.4]{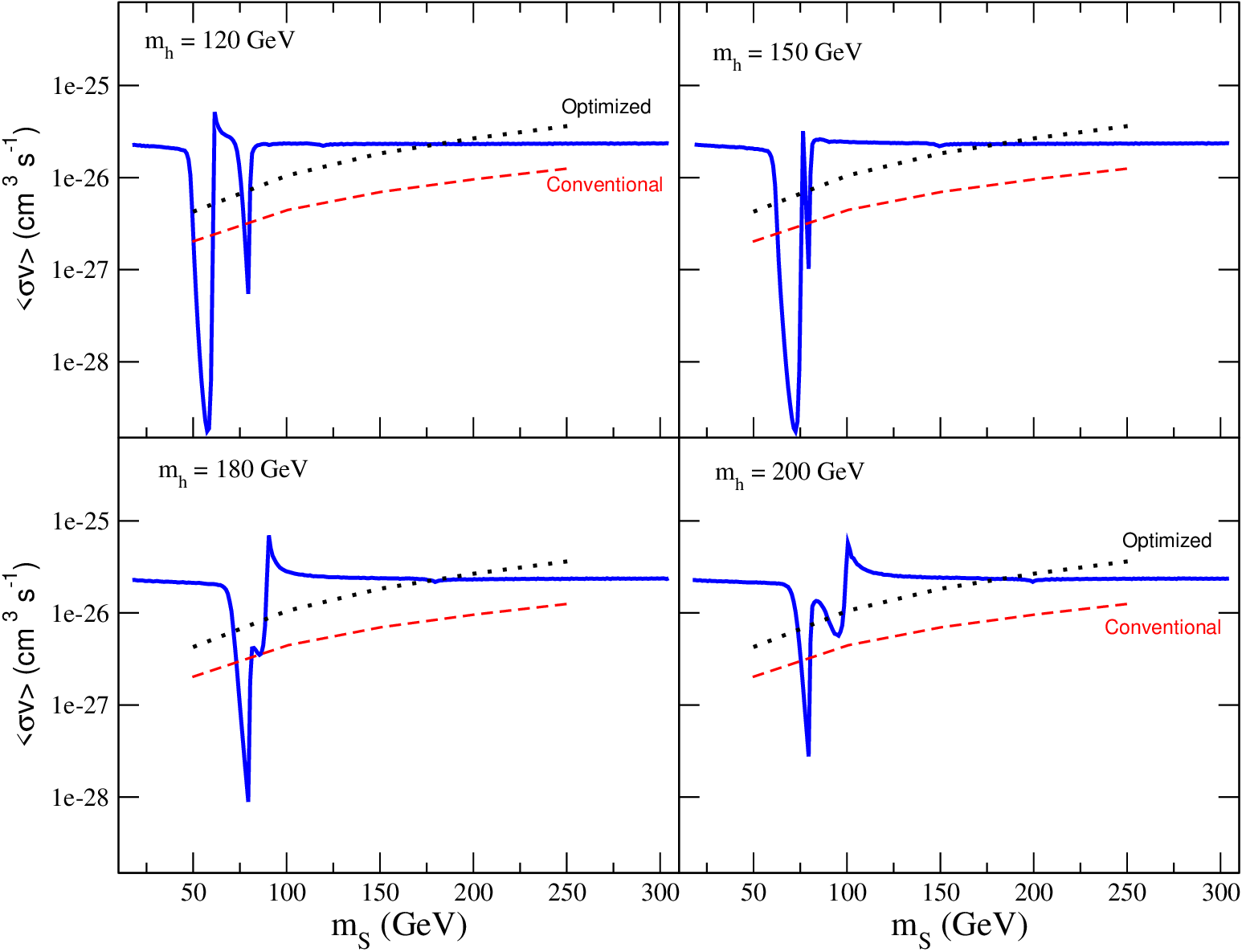}
\caption{The total annihilation cross section at low velocities as a function of $m_S$ for different values of $m_h$. The dashed (dotted) line shows the $\langle\sigma v\rangle$ required  to observe a DM annihilation signal at $3\sigma$ significance with one year of Fermi-GLAST data considering the conventional (optimized) diffuse model as background --see figure 9 in \cite{Baltz:2008wd}. \label{sigmav}}
\end{center}
\end{figure}

Even though the Galactic Center is expected to be the strongest source of $\gamma$ rays from DM annihilation, it is not necessarily the best place to observe them. To do that, one must be able to distinguish the DM signal from the Galactic diffuse background, a task that, given its large backgrounds,  might be more difficult at the Galactic Center. In fact, a recent study, \cite{Baltz:2008wd}, found that, for a NFW profile, the Fermi-GLAST sensitivity to a $\gamma$ ray signal from  the Galactic Halo (excluding the Center) is larger than that from the Galactic Center --compare figures $5$ and $9$ in \cite{Baltz:2008wd}. In the following, we will use those results, in particular figure $9$, to determine the Fermi-GLAST sensitivity to the singlet scalar model of dark matter.

In figure \ref{sigmav} we show $\langle\sigma v\rangle$, at small $v$, as a function of the singlet mass and different values of $m_h$. It is clearly seen from the figure that away from the higgs resonance and the $W$ threshold $\langle\sigma v\rangle$ is essentially constant and equal to the so-called typical annihilation cross section, $\langle\sigma v\rangle\sim 3\times 10^{-26}\mathrm{cm^3}\mathrm{s^{-1}}$. The dashed line shows the $\langle\sigma v\rangle$ required to make an observation of DM annihilation at $3\sigma$ significance with one year of FGST data considering a \emph{conventional} diffuse model \cite{Strong:1998fr} as background. The dotted lines shows the analogous quantity but using the \emph{optimized} diffuse model \cite{Strong:2004de} as background. These two lines were obtained by assuming that the region within $10^\circ$ of the GC is excluded from the analysis. Notice from the figure that for the optimized model most of region with $m_S\lesssim 175$ GeV is ``detectable by  FGST''. For the conventional diffuse model, the detectable region extends to much higher masses. It is fair to say, therefore, that Fermi-GLAST will probe a significant region of the parameter space of the singlet scalar model of dark matter.

\section{Conclusion}
We have studied in detail the singlet scalar model of dark matter. First, we introduced the model and presented an updated analysis of its parameter space. The dark matter constraint was found to reduce the viable parameter space to just two variables: the singlet mass and the higgs mass. Then, we computed the direct detection cross section and found that singlet masses below $50$ GeV are already ruled out by the recent data from CDMS and Xenon10.  Finally,  we calculated the expected gamma ray flux from the Galactic Center as well as the total annihilation cross section and showed that  a significant region of the parameter space  will be probed by Fermi-GLAST.  

\section*{Acknowledgments}
I am  supported by the \emph{Juan de la Cierva} program of the Ministerio de Educacion y Ciencia of Spain, by Proyecto Nacional FPA2006-01105, and by the Comunidad de Madrid under Proyecto HEPHACOS S-0505/ESP-0346. I also thank the ENTApP Network of the ILIAS project RII3-CT-2004-506222 and
the Universet Network MRTN-CT-2006-035863

\end{document}